\documentclass[aps,twocolumn,floatfix,showpacs,prb]{revtex4}
\usepackage{graphicx}
\usepackage{amsmath}
\usepackage{amssymb}
\usepackage{bm}

\begin{document}

\title{Coulomb stability of the $\bm{4\pi}$-periodic Josephson effect of Majorana fermions}
\author{B. van Heck}
\affiliation{Instituut-Lorentz, Universiteit Leiden, P.O. Box 9506, 2300 RA Leiden, The Netherlands}
\author{F. Hassler}
\affiliation{Instituut-Lorentz, Universiteit Leiden, P.O. Box 9506, 2300 RA Leiden, The Netherlands}
\author{A. R. Akhmerov}
\affiliation{Instituut-Lorentz, Universiteit Leiden, P.O. Box 9506, 2300 RA Leiden, The Netherlands}
\author{C. W. J. Beenakker}
\affiliation{Instituut-Lorentz, Universiteit Leiden, P.O. Box 9506, 2300 RA Leiden, The Netherlands}

\date{August, 2011}

\begin{abstract}
The Josephson energy of two superconducting islands containing Majorana fermions is a $4\pi$-periodic function of the superconducting phase difference. If the islands have a small capacitance, their ground state energy is governed by the competition of Josephson and charging energies. We calculate this ground state energy in a ring geometry, as a function of the flux $\Phi$ enclosed by the ring, and show that the dependence on the Aharonov-Bohm phase $2e\Phi/\hbar$ remains $4\pi$-periodic regardless of the ratio of charging and Josephson energies --- provided that the entire ring is in a topologically nontrivial state. If part of the ring is topologically trivial, then the charging energy induces quantum phase slips that restore the usual $2\pi$-periodicity.
\end{abstract}

\maketitle

The energy $H_{J}$ of a tunnel junction between two superconductors (a Josephson junction) depends on the difference $\phi$ of the phase of the order parameter on the two sides of the junction. The derivative $I_{J}=(2e/\hbar)dH_{J}/d\phi$ gives the supercurrent flowing through the junction in the absence of an applied voltage. In a ring geometry, the supercurrent depends periodically on the flux $\Phi$ enclosed by the ring, with periodicity $h/2e$. This familiar {\sc dc} Josephson effect\cite{Jos62,Tin04} acquires a new twist if the junction contains Majorana fermions.\cite{Kit01,Kwo03,Fu09}

Majorana fermions are charge-neutral quasiparticles bound to midgap states, at zero excitation energy, which appear in a socalled topologically nontrivial superconductor.\cite{Has10,Qi10} While in the conventional Josephson effect only Cooper pairs can tunnel (with probability $\tau\ll 1$), Majorana fermions enable the tunneling of single electrons (with a larger probability $\sqrt{\tau}$). The switch from $2e$ to $e$ as the unit of transferred charge amounts to a doubling of the fundamental periodicity of the Josephson energy, from $H_{J}\propto \cos\phi$ to $H_{J}\propto\cos(\phi/2)$. In a ring geometry, the period of the flux dependence of the supercurrent $I_{J}$ doubles from $2\pi$ to $4\pi$ as a function of the Aharonov-Bohm phase\cite{note2} $\varphi_{0}=2e\Phi/\hbar$. This $4\pi$-periodic Josephson effect has been extensively studied theoretically, \cite{Fu09,Lut10,Ore10,Ios11,Nog11,Law11,Jia11} as a way to detect the (so far, elusive) Majorana fermions.\cite{Ser11}

Since the Majorana fermions in a typical experiment will be confined to superconducting islands of small capacitance $C$, the Coulomb energy $H_{C}=Q^{2}/2C$ associated with a charge difference $2Q$ across the junction competes with the Josephson energy. The commutator $[\phi,Q]=2ei$ implies an uncertainty relation between charge and phase differences, so that a nonzero $H_{C}$ introduces quantum fluctuations of $\phi$ in the ground state.\cite{Tin04} What is the fate of the $4\pi$-periodic Josephson effect?

As we will show in this paper, the supercurrent through the ring remains a $4\pi$-periodic function of $\varphi_{0}$, regardless of the relative magnitude of $H_{C}$ and $H_{J}$. This Coulomb stability requires that all weak links in the ring contain Majorana fermions. If the ring has a topologically trivial segment, then quantum phase slips restore the conventional $2\pi$-periodicity of the Josephson effect on sufficiently long time scales. We calculate the limiting time scale for the destruction of the $4\pi$-periodic Josephson effect by quantum phase slips and find that it can be much shorter than the competing time scale for the destruction of the $4\pi$-periodicity by quasiparticle poisoning.\cite{Fu09}

\begin{figure}[tb] 
\centerline{\includegraphics[width=0.8\linewidth]{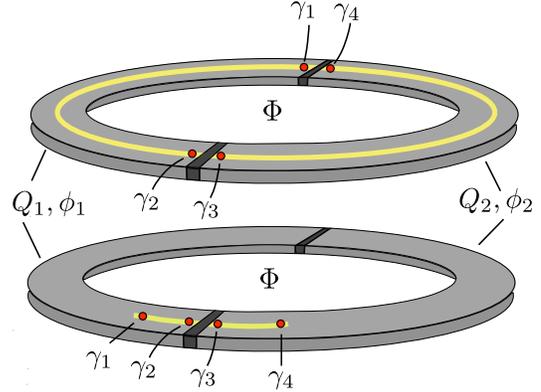}}
\caption{\label{fig_layout} 
Geometry of a {\sc dc squid}, consisting of a superconducting ring (grey) interrupted by two tunnel junctions (black) and threaded by a magnetic flux $\Phi$. A semiconductor nanowire (yellow) contains Majorana fermions at the end points (red dots). The two panels distinguish the cases that Majorana fermions are present at both junctions (top), or only at a single junction (bottom). The $4\pi$-periodic Josephson effect is stable against quantum phase slips in the first case, but not in the second case.
}
\end{figure}

We apply the general theory of Majorana-Josephson junction arrays of Xu and Fu\cite{Xu10} to the {\sc dc squid} geometry of Fig.\ \ref{fig_layout}, consisting of two superconducting islands separated by tunnel junctions. The islands have a charge difference $2Q=Q_{1}-Q_{2}$, with $Q_{n}=-2ei\partial/\partial\phi_{n}$ canonically conjugate to the superconducting phase $\phi_{n}$. The gauge invariant phase differences across the two junctions are given by $\phi=\phi_{1}-\phi_{2}$ and $\varphi_{0}-\phi$. Here we assume that the ring is sufficiently small that the flux generated by the supercurrent can be neglected, so the enclosed flux equals the externally applied flux.\cite{note1}

Each island contains a segment of a semiconductor nanowire, driven into a topologically nontrivial superconducting state by the proximity effect.\cite{Lut10,Ore10} (Alternatively, the nanowire could be replaced by the conducting edge of a two-dimensional topological insulator.\cite{Fu09}) The Majorana fermions appearing at the end points of each segment are represented by anti-commuting Hermitian operators $\gamma_{1},\gamma_{2},\gamma_{3},\gamma_{4}$ that square to unity,
\begin{equation}
\gamma_{n}=\gamma_{n}^{\dagger},\;\;\gamma_{n}\gamma_{m}+\gamma_{m}\gamma_{n}=2\delta_{nm}.\label{gammadef}
\end{equation}
The Majorana fermions are coupled by the tunnel junction. We distinguish two cases. In the first case (top panel in Fig.\ \ref{fig_layout}) each of the two tunnel junctions couples a pair of Majorana fermions. In the second case (bottom panel) one pair of Majorana fermions is coupled by a Josephson junction, while the other pair remains isolated. 

The Hamiltonian $H=H_{C}+H_{J,1}+H_{J,2}$ is the sum of charging and Josephson energies,
\begin{align}
&H_{C}=\frac{1}{2C}(Q+q_{\rm ind})^{2},\label{HCdef}\\
&H_{J,1}=E_{M,1}\Gamma_{1}\cos\frac{\phi}{2}-E_{J,1}\cos\phi,\label{HJ1def}\\
&H_{J,2}=E_{M,2}\Gamma_{2}\cos\frac{\varphi_{0}-\phi}{2}-E_{J,2}\cos(\varphi_{0}-\phi),\label{HJ2def}\\
&\Gamma_{1}=i\gamma_{2}\gamma_{3},\;\;
\Gamma_{2}=i\gamma_{4}\gamma_{1}.\label{Gammadef}
\end{align}
The induced charge $q_{\rm ind}=C_{g}V_{g}$ accounts for charges on nearby electrodes, controlled by a gate capacitance $C_{g}$ and gate voltage $V_{g}$. The energy scales $E_{M,n}$ and $E_{J,n}$ quantify the Josephson coupling strength of, respectively, single electrons and electron pairs. With this Hamiltonian we can describe both cases considered, by putting $E_{M,2}=0$ for the junction without Majorana fermions.

The eigenstates $\Psi(\phi_{1},\phi_{2})$ of $H$ should satisfy the fermion parity constraint \cite{Fu10}
\begin{align}
&\Psi(\phi_{1}+2\pi n,\phi_{2}+2\pi m)=(-1)^{nq_{1}}(-1)^{mq_{2}}\Psi(\phi_{1},\phi_{2}),\label{parity1}\\
&q_{n}=\tfrac{1}{2}(1-p_{n}),\;\;p_{1}=i\gamma_{1}\gamma_{2},\;\;p_{2}=i\gamma_{3}\gamma_{4}.\label{parity2}
\end{align}
The operators $q_{n}$ and $p_{n}$ have, respectively, eigenvalues $0,1$ and $\pm 1$, depending on whether island $n$ contains an even or an odd number of electrons. The constraint \eqref{parity1} enforces that the eigenvalues of $Q_{n}$ are even multiples of $e$ for $q_{n}=0,p_{n}=1$ and odd multiples of $e$ for $q_{n}=1,p_{n}=-1$.

It is possible to solve the eigenvalue problem $H\Psi=E\Psi$ subject to the constraint \eqref{parity1}, along the lines of Ref.\ \onlinecite{Xu10}, but alternatively one can work in an unrestricted Hilbert space. The restriction is removed by the unitary transformation
\begin{equation}
\Psi=U_{1}U_{2}\tilde{\Psi},\;\;U_{n}=\exp(iq_{n}\phi_{n}/2).\label{Udef}
\end{equation}
The function $\tilde{\Psi}(\phi_{1},\phi_{2})$ is $2\pi$-periodic in each of its arguments, so the constraint \eqref{parity1} is automatically satisfied. Now the eigenvalues of $Q_n$ are all even multiples of $e$. The transformed Hamiltonian $\tilde{H}=(U_{1}U_{2})^{\dagger}HU_{1}U_{2}$ becomes
\begin{align}
&\tilde{H}=\frac{1}{2C}\Bigl(Q+\frac{eq_{1}-eq_{2}}2+q_{\rm ind}
  \Bigr)^{2}\nonumber\\
&\quad+\tfrac{1}{2}\bigr[e^{-iq_{1}\phi_{1}}\bigl(E_{M,1}\Gamma_{1}+E_{M,2}\Gamma_{2} e^{i\varphi_{0}/2}\bigr)e^{iq_{2}\phi_{2}}+\text{H.c.}\bigr]\nonumber\\
&\quad-E_{J,1}\cos\phi-E_{J,2}\cos(\varphi_{0}-\phi),\label{Hprimedef}
\end{align}
where we have used the identity
\begin{equation}
U_{n}^{\dagger}\Gamma_{m} e^{i\phi_{n}/2}=\Gamma_{m} U_{n}.\label{UGammaidentity}
\end{equation}
Notice that the Hamiltonian has become $2\pi$-periodic in the superconducting phases $\phi_{1},\phi_{2}$, while remaining $4\pi$-periodic in the flux $\varphi_{0}$. Notice also that $\tilde{H}$ may depend on the $\phi_{n}$'s separately, not just on their difference. This does not violate charge conservation, because the conjugate variables $Q_{n}$ now count only the number of Cooper pairs on each island --- not the total number of electrons.

The four Majorana fermions encode a qubit degree of freedom.\cite{Nay08} The states of the qubit are distinguished by the parity of the number of electrons on each island. If the total number of electrons in the system is even (${\cal P}=1$), the qubit states are $|11\rangle$ and $|00\rangle$, while for an odd total number of electrons (${\cal P}=-1$) the states are $|10\rangle$ and $|01\rangle$. In this qubit basis, the products of Majorana operators appearing in the Hamiltonian \eqref{Hprimedef} are represented by Pauli matrices,
\begin{equation}
q_{1}=\tfrac{1}{2}+\tfrac{1}{2}\sigma_{z},\;\;q_{2}=\tfrac{1}{2}+\tfrac{1}{2}{\cal P}\sigma_{z},\;\;
\Gamma_{1}=-\sigma_x,\;\;\Gamma_{2}={\cal P}\sigma_x.\label{gGammasigma}
\end{equation}

It is straightforward to calculate the eigenvalues of $\tilde{H}$, by evaluating its matrix elements in the basis of eigenstates of $Q$. The spectrum $E_{n}^{\cal P}(\varphi_{0},q_{\rm ind})$ as a function of the enclosed flux and the induced charge has two branches distinguished by the total fermion parity ${\cal P}=\pm 1$, with
\begin{equation}
E_{n}^{+}(\varphi_{0},q_{\rm ind})=E_{n}^{-}(\varphi_{0}+2\pi,q_{\rm ind}+e/2).\label{EPsymmetry} 
\end{equation}
We first consider the case that both junctions contain Majorana fermions (top panel in Fig.\ \ref{fig_layout}).

\begin{figure}[tb]
\centerline{\includegraphics[width=0.8\linewidth]{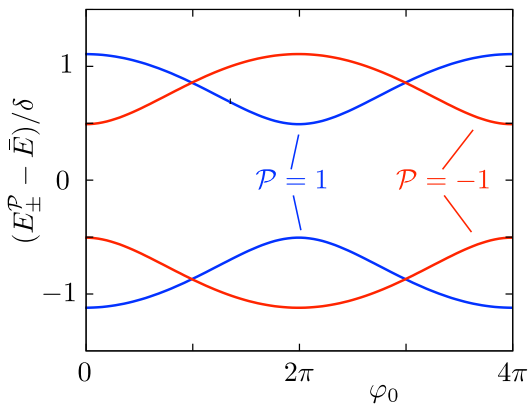}}
\caption{\label{fig_spectrum}
Spectrum of the {\sc dc squid} in the top panel of Fig.\ \ref{fig_layout}, containing Majorana fermions at both Josephson junctions. The curves are the result \eqref{large_charge}, in the limit that the charging energy dominates over the Josephson energy. The parameters chosen are $E_{M,1}=E_{M,2}=\delta$. The level crossing is between states of different fermion parity ${\cal P}$, and therefore there can be no tunnel splitting due to the Coulomb interaction (which conserves ${\cal P}$).
}
\end{figure}

A fully analytical calculation is possible in the limit that the charging energy dominates over the Josephson energy ($E_{C}\equiv e^{2}/2C\gg E_{M,n},E_{J,n}$). Only the two eigenstates of $Q$ with lowest charging energy $\bar{E}\pm\frac{1}{2}\delta$ are needed in this limit and $2e$ tunnel processes may be neglected relative to $e$ tunnel processes (so we may set $E_{J,n}=0$). We thus obtain the simple expression
\begin{equation}
 E_{\pm}^{\cal P}= \bar{E}
 \pm \tfrac{1}{2} \biggl[ \delta^2+E_{M,1}^{2}+E_{M,2}^{2} +  2\mathcal{P} E_{M,1}E_{M,2}
 \cos \frac{\varphi_0}2 \biggr]^{1/2}.\label{large_charge}
\end{equation}
The resulting $4\pi$-periodic spectrum is shown in Fig.\ \ref{fig_spectrum}. 

The crossing of the two branches $E_{-}^{+}$ and $E_{-}^{-}$ at $\varphi_{0}=\pi$ is protected, regardless of the value of $E_{C}$, because the charging energy cannot couple states of different ${\cal P}$. Quasiparticle poisoning (the injection of unpaired electrons) switches the fermion parity on a time scale $T_{p}$, which means that the $4\pi$-periodicity of the energy of the ring can be observed if the enclosed flux is increased by a flux quantum in a time $T_{\Phi}\ll T_{p}$. 

\begin{figure}[tb]
\centerline{\includegraphics[width=0.8\linewidth]{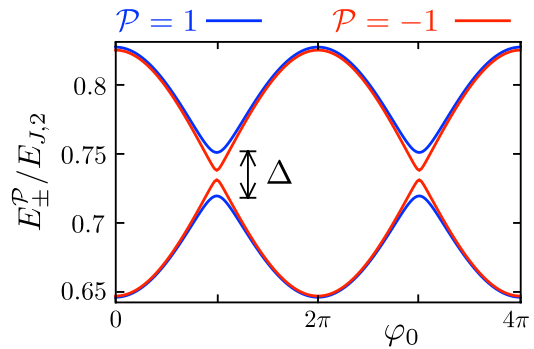}}
\caption{\label{fig_numerics}
Spectrum of the {\sc dc squid} in the bottom panel of Fig.\ \ref{fig_layout}, containing Majorana fermions at only one of the two Josephson junctions. The curves are a numerical calculation for the full Hamiltonian, in the regime that the Josephson energy of the trivial junction is the largest energy scale. The parameters chosen are $E_{J,2}=4E_{C}=10E_{M,1}$, $E_{M,2}=0=E_{J,1}$, and $q_{\rm ind}=0$. In contrast to Fig.\ \ref{fig_spectrum}, a tunnel splitting $\Delta$ appears because the level crossing is between states of \textit{the same} fermion parity.
}
\end{figure}

We now turn to the case that one of the two Josephson junctions does not contain Majorana fermions (lower panel in Fig.\ \ref{fig_layout}). By putting $E_{M,2}=0$ the Hamiltonian becomes $2\pi$-periodic in $\varphi_{0}$. In Fig.\ \ref{fig_numerics} we show the spectrum for a relatively large Josephson energy of the trivial junction. The phase $\phi$ is then a nearly classical variable, which in the ground state is close to $\varphi_{0}$ (mod $2\pi$). The charging energy opens a gap in the spectrum near $\varphi_{0}=\pi$ (mod $2\pi$), by inducing tunnel processes from $\phi=\varphi_{0}$ to $\phi=\varphi_{0}\pm2\pi$ (quantum phase slips). A tunnel splitting by the ${\cal P}$-conserving charging energy is now allowed because the level crossing is between states of the same ${\cal P}$.

\begin{figure}[tb]
\centerline{\includegraphics[width=0.8\linewidth]{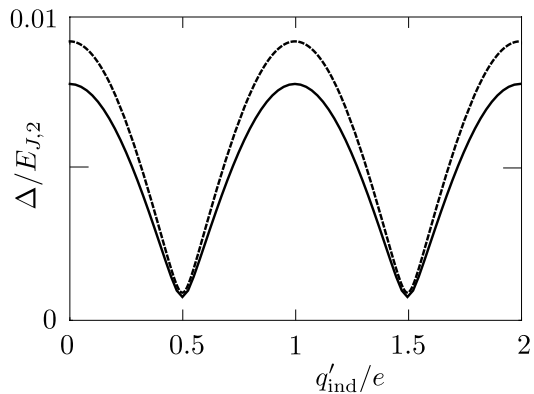}}
\caption{\label{fig_splitting}
Tunnel splitting at $\varphi_0=\pi$ as a function of the induced charge. The dashed curve correspond to Eq.\ \eqref{eq:tunnel}, the solid curve to numerical calculations for the full Hamiltonian, for $E_{J,2}=5\,E_C=25\,E_{M,1}$ (with $E_{M,2}=0=E_{J,1}$). 
}
\end{figure}

A semiclassical calculation of the tunnel splitting due to quantum phase slips at the trivial Josephson junction, along the lines of Ref.\ \onlinecite{Koc07}, gives for $E_{J}\equiv E_{J,2}\gg E_{C}\gg E_{M,1}\equiv E_{M}$ the spectrum
\begin{align}
&E^{\cal P}_\pm=-E_{J}+ \sqrt{2 E_C E_{J}} \pm \sqrt{ E_{M}^{2}
  \cos^2 (\varphi_0/2) + \Delta^2},\label{eq:gst_energy2}\\
&  \Delta =16 \bigl(E_{C} E_{J}^{3}/2\pi^{2}\bigr)^{1/4}\exp\bigl(-\sqrt{8 E_{J}/E_C}\bigr)\nonumber\\
&\quad\quad\mbox{}\times  \sqrt{\cos^{2}(\pi q'_{\rm ind}/e)+\frac{\pi^{2}E_{M}^{2}}{8E_{C}E_{J}}\sin^{2}(\pi q'_{\rm ind}/e)},\label{eq:tunnel}
\end{align}
where we have abbreviated $q'_{\rm ind}=q_{\rm ind}+(e/4)(1-{\cal P})$. The second term on the right-hand-side of Eq.\ \eqref{eq:gst_energy2} describes the
effect of zero-point fluctuations of $\phi$ around the values $\varphi_{0}$ and $\varphi_{0}\pm 2\pi$. Tunnel processes $\phi=\varphi_{0}\mapsto\varphi_{0}+2\pi$ and
$\phi=\varphi_{0}\mapsto\varphi_{0}-2\pi$ produce the third term. The sine and cosine factors in Eq.\ \eqref{eq:tunnel} accounts for interference between these two quantum phase slip processes (Aharonov-Casher effect).\cite{Fri02,Tiw07,Has10,Gro11a,Gro11b} The numerical calculation in Fig.\ \ref{fig_splitting} agrees quite well with the semiclassical approximation \eqref{eq:tunnel}.

The tunnel splitting $\Delta$ ensures that the energy of the ring evolves $2\pi$-periodically if the flux $\Phi$ is increased by a flux quantum $h/2e$ in a time $T_{\Phi}$ which is long compared to $T_{\Delta}=\hbar E_{M,1}/\Delta^{2}$. For $T_{\Phi}\lesssim T_{\Delta}$ there is a significant probability $\exp(-T_{\Phi}/T_{\Delta})$ for a Landau-Zener transition through the gap, resulting in a $4\pi$-periodic evolution of the energy. 

This limiting time scale $T_{\Delta}$ originating from quantum phase slips can be compared with the time scale $T_{p}$ for quasiparticle poisoning. We require $T_{\Phi}$ small compared to both $T_{\Delta}$ and $T_{p}$ to observe the $4\pi$-periodic Josephson effect. For $\Delta> (\hbar E_{M,1}/ T_{p})^{1/2}$ one has $T_{\Delta}<T_{p}$, so quantum phase slips govern. A recent experiment finds $T_{p}\simeq 2\,{\rm ms}$ in Al for temperatures below $160\,{\rm mK}$.\cite{Vis11} Since $E_{M,1}$ will be well below $1\,{\rm meV}$, one has $T_{\Delta}<T_{p}$ if quantum phase slips occur with a rate $\Delta/\hbar$ higher than $30\,{\rm MHz}$. While quantum phase slip rates can vary over many orders of magnitude due to the exponent in Eq.\ \eqref{eq:tunnel}, typical values for a {\sc dc squid} are in the GHz range.

In conclusion, we have shown that Coulomb charging effects do not spoil the $4\pi$-periodic Josephson effect in a superconducting ring, provided that all weak links contain Majorana fermions. Quantum phase slips at a weak link without Majorana fermions restore the $2\pi$-periodicity on time scales long compared to a time $T_{\Delta}$, which may well be shorter than the time scale for quasiparticle poisoning.

The origin of the protection of the $4\pi$ periodicity if the entire ring is topologically nontrivial is conservation of fermion parity.\cite{Fu09} (See Ref.\ \onlinecite{Ryu10} for a more general perspective.) This protection breaks down if part of the ring is a trivial superconductor, because then the level crossing involves states of the same fermion parity and tunnel splitting by the charging energy is allowed (see Fig.\ \ref{fig_numerics}). 

We note in closing that the different stability of the $4\pi$-periodic Josephson effect in the two geometries of Fig.\ \ref{fig_layout}, examined here with respect to Coulomb charging, extends to other parity-preserving perturbations of the Hamiltonian. For example, overlap of the wave functions of two Majorana bound states on the same island introduces a term $H_{\rm overlap}=i\epsilon\gamma_{1}\gamma_{2}$. For the lower panel of Fig.\ \ref{fig_layout}, this term leads to a tunnel splitting $\Delta=2\epsilon$ which spoils the $4\pi$-periodicity.\cite{Kit01} For the upper panel of Fig.\ \ref{fig_layout}, $\Delta\equiv 0$ because $H_{\rm overlap}$ preserves fermion parity.

This research was supported by the Dutch Science Foundation NWO/FOM and by an ERC Advanced Investigator Grant. We have learned of independent work on a related problem by L. Fu and thank him for valuable discussions.


\begin{thebibliography}{99}
\bibitem{note2} As a function of the enclosed flux, $I_{J}$ has the same $h/e$ periodicity as the persistent current $I_{N}$ through a normal metal ring (radius $R$). One can distinguish the two currents by their size dependence: While $I_{N}$ decays as $1/R$ or faster, $I_{J}$ has the $R$-independence of a supercurrent.
\bibitem{note1} The flux induced by the supercurrent $I_{J}$ due to the nonzero inductance $L\simeq \mu_{0}R$ of the ring may be neglected relative to the applied flux if $L I_{J} \ll \hbar/e$. The magnitude of the supercurrent can be estimated by $\hbar I_{J}/e\simeq \min (E_{J}, E_{J}^{2}/E_{C})\equiv E_{c}$. For $E_c \simeq 1\,{\rm meV}$ the induced flux can be neglected if $R\ll 1\,{\rm cm}$.
\end{thebibliography}
\end{document}